\documentstyle[preprint,aps]{revtex}

  \newcommand{\mean}[1]{\mbox{$\left< #1  \right>$}}

  \newcommand{\cH}{\tilde{\cal H}_i}

  \title{Dissipative quantum systems modeled by a two level
reservoir coupling}
  \author{A. O. Caldeira$^*$, A. H. Castro Neto$^{\dag}$ and T.
Oliveira de Carvalho$^*$
\begin{center}
 $^*${\small \it Departamento de F\'{\i}sica do Estado S\'olido e
Ci\^encia dos Materiais,} \\
  {\small \it Instituto de F\'{\i}sica Gleb Wataghin, CP 6165,} \\
{\small \it Universidade Estadual de Campinas
13081-970 \, \ Campinas, S\~ao Paulo, Brazil\\}
\vspace{1cm}
$^{\dag}${\small \it Loomis Laboratory of Physics, University
of Illinois at} \\
{\small \it Urbana-Champaign, 1110 West Green Street, Urbana, Illinois, 61801}
\end{center}}

\begin{document}

  \maketitle

  \begin{abstract}
  The coupling between a quantum dynamical system and a two-level
system reservoir is analysed within the framework of the
Feynman-Vernon theory. We stress the differences between this new
reservoir and the well-known bath of oscillators and show that, in
order to obtain the Langevin equation for the system of interest in
the high temperature regime, we have to choose a spectral
distribution function $J(\omega)$ which is finite for $\omega=0$.
   \end{abstract}

\newpage

In the two past decades, there has been great effort in bringing out
a theoretical treatment of dissipation in quantum systems whose
dynamics is described by the Langevin equation
\begin{equation}
m\ddot{x} + \eta \dot{x} +\frac{\partial V}{\partial x} = F(t)
\end{equation}
where $F(t)$ is a random force:
\begin{equation}
\left< F(t) \right> =0
\end{equation}

One possible way to achieve that is by coupling the quantum system of
interest to a reservoir \cite{verno}. Under very general conditions
\cite{calde1,calde2} we can choose the reservoir to be a set of
decoupled harmonic oscillators, each of which has a different
characteristic frequency $\omega_k$. These frequencies are distributed
in such a way that the spectral function $J(\omega)$ is given by
\begin{equation}
J(\omega)= \left\{
\begin{array}{cl}
\eta \omega  & \mbox{for $\omega <
\Omega$, and} \\
0 & \mbox{for $\omega > \Omega$}
\end{array}
\right.
\end{equation}
in order that the two time correlation of the random force becomes
\begin{equation}
\left< F(t)F(t') \right> = \frac{1}{2 \pi} \int e^{-i\omega (t-t')}
\, \eta \hbar \omega \coth \left(\frac{\beta \hbar \omega}{2}\right)
\, d\omega \ ,
\end{equation}
where $\beta=\frac{1}{kT}$ and a cut-off frequency for $\omega$ is
assumed, say $\Omega$. In the long time and high temperature limit:
\begin{eqnarray}
|t-t'| \gg \Omega^{-1}  & \mbox{ and } & \beta \gg \hbar |t-t'|^{-1}
\end{eqnarray}
one obtains
\begin{equation}
\mean{F(t)F(t')} = 2\eta k T \delta (t-t') \ ,
\end{equation}
which is the classical two time correlation of the random force.

We propose here another kind of reservoir: a set of decoupled two
level systems (which may be understood as a projection onto the first
two levels of the bath of oscilators) which is coupled to the system
of interest by
\begin{equation}
{\cal H} = {\cal H}_S + {\cal H}_i + {\cal H}_b
\end{equation}
where ${\cal H}_S$ represents the Hamiltonian of the system we are
interested in and
\begin{eqnarray}
{\cal H}_i &=& -\sum_{k=1}^N J_k X \sigma_{x_k} \\
{\cal H}_b &=& \sum_{k=1}^N \frac{\hbar \omega_k}{2} \sigma_{z_k}
\end{eqnarray}

We assume that the system of interest and the bath are decoupled at
$t=0$, so that the initial condition is given by
\[ \rho(0) = e^{\beta{\cal H}_b} \tilde{\rho}(0) \]
as usual. We treat the interaction problem
within the Feynman-Vernon theory, obtaining for the influence
functional:
\begin{equation}
{\cal F}[x,y] = Tr_b \left[ e^{-\beta {\cal H}_b}
A_{y'y}(0,t)A_{xx'}(t,0) \right] \ ,
\label{trace}
\end{equation}
where we have defined
\begin{eqnarray}
A_{y'y}(0,t) &\equiv &  {\it T} exp \left[ +\frac{i}{\hbar} \int_0^t
{\tilde{\cal H}}_i(x(\tau))d\tau \right] e^{+i{\cal H}_b t/\hbar} \\
\mbox{and }\hspace{.2in} & & \nonumber \\
A_{xx'}(t,0) &\equiv & e^{-i{\cal H}_b t/\hbar} {\it T} \, exp \left[
-\frac{i}{\hbar} \int_0^t {\tilde{\cal H}}_i(x(\tau))d\tau \right]
\ .
\end{eqnarray}
where $T$ denotes the time ordered product and
\[ {\tilde{\cal H}_i} \equiv {\displaystyle e}^{\displaystyle i
{\cal H}_b t/\hbar} {\cal H}_i(x(\tau)) {\displaystyle
e}^{\displaystyle -i {\cal H}_b t/\hbar}  \]
is the interaction Hamiltonian in the interaction picture.

We calculate $A_{y'y}$ and $A_{xx'}$ up to second order in the
interaction strength $J_k$, see \cite{calde3}:
\begin{eqnarray}
A_{y'y}(0,t) &\approx &  \left\{ 1+
\frac{i}{\hbar} \int_0^t d\tau {\tilde{\cal H}_i}(y(\tau)) - \right.
\nonumber \\
& & \hspace{-2cm} \left. -\frac{1}{{\hbar}^2} \int_0^t d\tau \,
\int_0^{\tau} d\sigma  {\tilde{\cal H}}_i (y(\tau)){\tilde{\cal H}}_i
(y(\sigma)) \right\} e^{i{\cal H}_b t/\hbar} \\
\mbox{and } \hspace{.2in} & & \nonumber \\
A_{xx'}(t,0) &\approx &  e^{-i{\cal H}_b t/\hbar} \;
\left\{ 1+ \frac{i}{\hbar} \int_0^t d\tau {\tilde{\cal H}_i}(x(\tau)) -
\right.
\nonumber
\\ & & \left. -\frac{1}{{\hbar}^2} \int_0^t d\tau \,\int_0^{\tau}
d\sigma  {\tilde{\cal H}}_i (x(\tau)){\tilde{\cal H}}_i (x(\sigma))
\right\} \ .
\end{eqnarray}
We evaluate the trace (see (\ref{trace})) keeping the terms up to
second order in ${\cal H}_i$, and this leads to
\begin{eqnarray}
F[x,y] & \approx &  \nonumber \\
& & \hspace{-2.7cm}1 -\frac{1}{{\hbar}^2} \int_0^t d\tau \,
\int_0^{\tau} d\sigma \left\{ \mean{\cH(y(\sigma))\cH(y(\tau))}
 + \mean{\cH(x(\tau))\cH(x(\sigma))} \right. - \nonumber \\
& & \hspace{-1cm} -\mean{\cH(y(\tau))\cH(x(\sigma)) }\left. -
\mean{\cH(y(\sigma))\cH(x(\tau))}\right\} \ ,
\label{if1}
\end{eqnarray}
where the linear terms in $\mean{\cH}$ vanishes, since $\cH$ has only
zeros in its diagonal.

We can carry out the averages $\mean{\;\;}$ over the reservoir
obtaining for the influence functional\cite{foot}
\begin{eqnarray}
F[q,\xi]
= &\exp & \left\{ - \frac{1}{\hbar^2} \int_0^t d\tau \,
\int_0^{\tau} d\sigma \sum_{\alpha} {J}_{\alpha}^2 \left[
\xi(\tau)\xi(\sigma) \cos{\omega_{\alpha} (\tau - \sigma)}
-\right.\right. \nonumber \\
& & \;\;\;\;\left.\left. \;\;- 2i q(\tau)\xi(\sigma)
\tanh{\frac{\beta \hbar \omega_{\alpha}}{2}}
\sin{\omega_{\alpha}(\tau-\sigma)} \right] \right\} \ ,
\end{eqnarray}
where
\begin{eqnarray*}
q &=& \frac{x+y}{2} \\
\xi &=& x- y
\end{eqnarray*}

Taking a continuous distribution of two level systems, we define as
in \cite{calde2},
\begin{equation}
J(\omega) = \sum_\alpha \frac{J_{\alpha}^2}{\hbar} \delta(\omega
-\omega_{\alpha})
\end{equation}
and write the final form of the influence functional for the system
of interest as
\begin{eqnarray}
F[q,\xi] &=& exp -\frac{1}{\hbar} \int_0^t d\tau \, \int_0^{\tau}
d\sigma \, \int_0^{\infty} d\omega \, J(\omega) \left[
\xi(\tau)\xi(\sigma) \cos{\omega(\tau -\sigma)} \right. - \nonumber \\
& & \hspace{2cm} \left. - 2i q(\tau)\xi(\sigma) \tanh{\frac{\beta \hbar
\omega}{2}}\sin{\omega(\tau-\sigma)} \right] \ .
\label{orda}
\end{eqnarray}

Now we may compare this influence functional with that obtained in
\cite{calde2}, for the bath of oscilators. If we take the same
spectral function as in \cite{calde1,calde2}, that means
$J(\omega)=\eta \omega\,\theta(\Omega-\omega)$, this new environment
gives us a very different behaviour from the previous model. As we
easily see from (\ref{orda}) the imaginary part of the exponent of the
influence functional becomes explicitly temperature dependent and its
frequency dependence will not allow us to write a `memoriless'
damping term as before. This only happens at very low temperatures
($\beta\rightarrow \infty$) when the damping term turns out to
be exactly as before.

On the other hand, the real part of the exponent of the influence
functional has no temperature dependence at all. The form of that
term clearly show that we cannot ever write a diffusion coefficient
for this case. This term will always give a non-Markoffian diffusion
process.

In order to recover the same behaviour, as with the bath of oscillators, onone
should work with a new
$J(\omega)$ given by
\begin{equation}
J(\omega)= \left\{
\begin{array}{cl}
\eta \omega \coth{\frac{\beta \hbar \omega}{2}}  & \mbox{for
$\omega < \Omega$, and} \\
0 & \mbox{if $\omega > \Omega$}
\end{array}
\right.
\label{concc}
\end{equation}
which is finite for $\omega \rightarrow 0$ and temperature dependent.

Although at the present stage our results are purely academical, we
think this could be a reasonable starting point for dealing with the
relaxation of the magnetization of a spin glass \cite{toc5}.

T.O.C. wishes to acknowledge the support of the FAPESP and the FAEP
(Funda\-\c{c}\~ao de Amparo \`a Pesquisa do Estado de S\~ao Paulo and
Funda\c{c}\~ao de Apoio ao Ensino e \`a Pesquisa--Unicamp). A.H.C.N.
and A.O.C. acknowledge the CNPq (Conselho Nacional de Desenvolvimento
Cient\'{\i}fico e Tecnol\'ogico) for a grant and partial support
respectively.


\begin{thebibliography}{1}
\bibitem{foot}
We have used the relation $1-x^2 = e^{-x^2}$, since we are
considering only terms up to second order in the interaction
Hamiltonian.
\end{thebibliography}
\end{document}